## Engineering of Dual Wavelength, Polarization Selective Metalenses in Silicon Carbide

Xiaoying Huang [a,b], Ziwei Yang [c], Konosuke Shimazaki [a,b], Kritsana Saego [a,b], Otto Cranwell Schaeper [a,b], Evan Williams [a,b], Dragomir Neshev [c], Hark Hoe Tan [c], Igor Aharonovich[a,b], and Mehran Kianinia* [a,b]

[a] School of Mathematical and Physical Sciences, University of Technology Sydney, Ultimo, New South Wales 2007, Australia

[b] ARC Centre of Excellence for Transformative Meta-Optical Systems, University of Technology Sydney, Ultimo, New South Wales 2007, Australia

[c] *Australian Research Council Centre of Excellence for Transformative Meta-Optical Systems, Department of Electronic Materials Engineering, Research School of Physics, The Australian National University, Canberra, ACT 2600, Australia.*

## Abstract

Spin defects in silicon carbide (SiC) are promising candidates for integrated quantum photonics, offering long-lived spin states and near-infrared emission suitable for low-loss photonic integration and fibre-based quantum communication. However, light extraction from these defects remains challenging due the relatively high refractive index of SiC. Metalenses offer a compact approach to enhance light collection by engineering the wavefront directly at the material interface. Here, we design and fabricate monolithic metalenses from SiC bulk material that simultaneously operate at 860 and 1240 nm, matching with emission from the nitrogen vacancy and silicon vacancy colour centers. By independently engineering the phase response at both wavelengths, the metalens enables collection and polarization manipulation of the emitted light. We further employ the metalenses to demonstrate optically detected magnetic resonance of both defects simultaneously. These multifunctional metalenses provide a compact optical interface for scalable integrated SiC photonic devices.

## Introduction

Silicon carbide (SiC) has established as one of the leading solid-state platforms for quantum technologies owing to its wide bandgap, mature semiconductor manufacturing platform, and rich family of optically active defects [1-3].In particular, silicon vacancy ($V_{Si}$) and nitrogen-vacancy (NV) defects in SiC, which emit in the near-infrared and infrared spectral regions, have attracted significant attention as promising quantum light sources [4,5]. These defects exhibit long spin coherence time and enable optical spin initialization, coherent manipulation, and readout, making them attractive candidates for quantum communication, quantum sensing, and quantum information processing [5-7]. Together with its excellent optical, thermal, and mechanical properties [8,9], SiC offers a versatile platform for scalable integrated quantum photonics.

Realizing the full potential of these colour centers, however, requires efficient optical interfaces capable of extracting and manipulating the emitted photons [10]. This remains a major challenge due to the high refractive index of SiC, which results in strong total internal reflection, limiting the extraction of emitted photons into free space [11,12]. To address this challenge, various photonic structures including nanopillars [13], solid immersion lenses [14], optical cavities [15] and waveguide integrated architectures [16] have been explored to enhance collection efficiency. However, developing integrated optical interfaces that simultaneously collect, shape, and manipulate photons at the source remains a critical step towards practical SiC based quantum photonic devices [17,18].

To this end, metasurfaces based on arrays of subwavelength nanostructures provide a promising route towards compact and multifunctional optical interfaces [18,19]. By engineering the phase, amplitude, and polarization response of individual nano-elements, metalenses can integrate multiple optical functionalities, including focusing, beam shaping, polarization control, and multiplexing, within a single planar device [20-22]. Recent demonstrations have combined metalenses with diamond colour centers, while monolithic SiC metalenses have established the feasibility of integrating meta-optical elements directly onto the SiC platform [23,24]. While demonstration of integrated defects in silicon carbide with monolithic metagenes remain elusive [25,26], existing metasurface devices are exclusively optimized for a single colour center operating at a single wavelength [27,28]. With emitters spanning a broad spectral range from 800 nm to 1300 nm for Vsi and NV centers in SiC, implementing multifunctional metasurface would improve alignment complexity and system integration challenges. Therefore, Integrated optical interfaces capable of simultaneously addressing spectrally distinct colour centers are therefore highly desirable

In this work, we design and fabricate a monolithic SiC metalens that simultaneously operates at 860 nm and 1240 nm with polarization-dependent optical functionality. We demonstrate the performance of the device by implanting SiC colour centers, including $V_{si}$ and NV spin defects, and efficiently collecting their emission through the fabricated metalens. By integrating wavelength-selective phase control within a single metasurface, our approach enables a compact and multifunctional optical interface for multiple SiC colour centers, providing a pathway towards scalable quantum photonic devices.

## Discussion

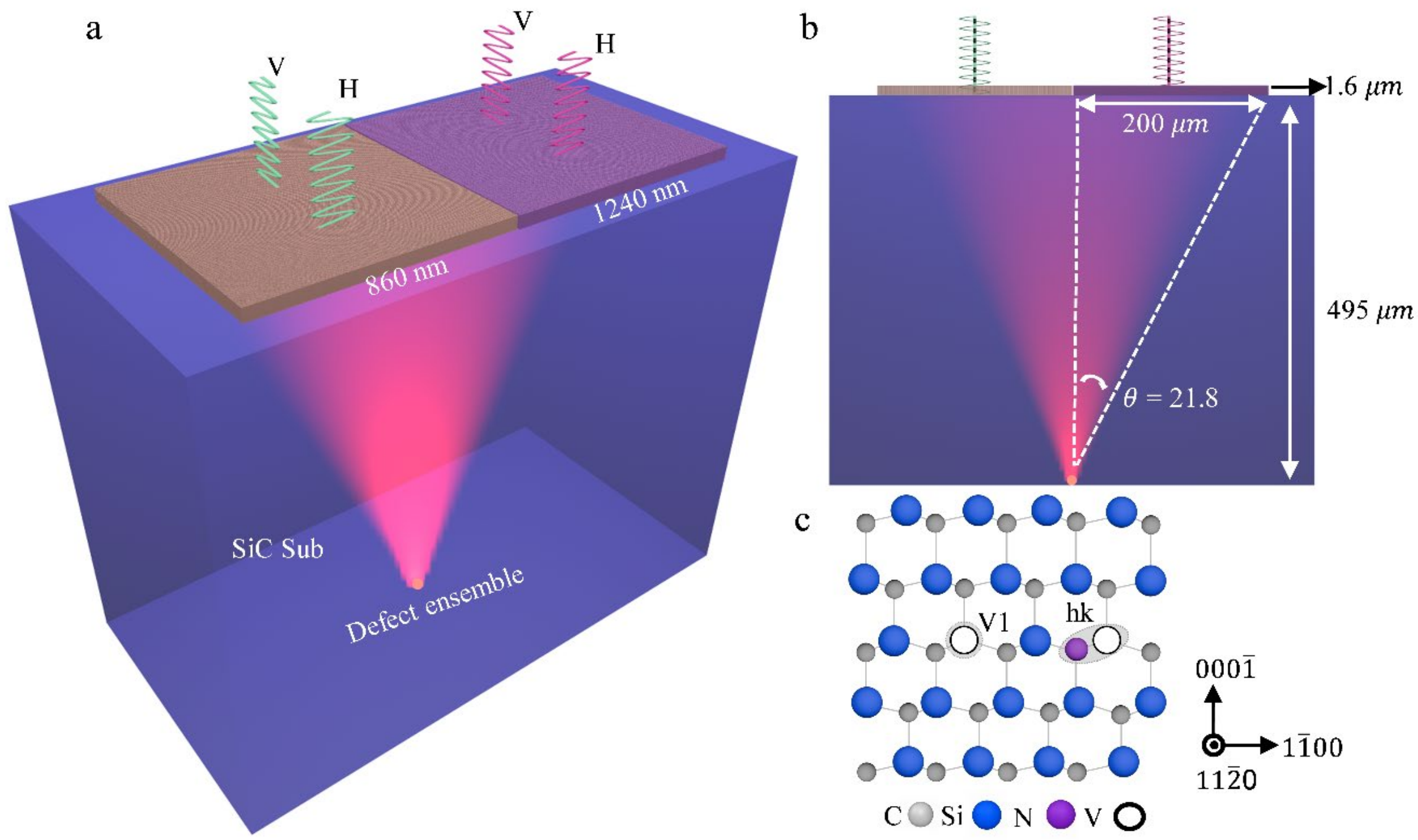

*Figure 1. Design of the monolithically integrated dual-wavelength SiC metalens. (a) A prospective of the monolithic metalens integrated on the top surface of a bulk SiC substrate. The design is to collect the defect ensemble emission from the bottom and realize wavelengths dependent with orthogonal V and H polarized light at 860 nm and 1240 nm, respectively. (b) A side-view of the device. it consists of two adjacent 200 * 200 µm parts designed for 860 nm (left) and 1240 nm (right), respectively. The metalens with a height of 1.6 µm height and a focal length of 495 µm to collect fluorescence from a defect ensemble located on the opposite side of the SiC substrate, corresponding to a collection half-angle of 21.8°. (c) Illustration of the defect ensemble crystal structure of the Silicon vacancy (V1) and Nitrogen vacancy (hk) defects along [1120] crystallographic direction of 4H-SiC.*

The schematic design of the monolithically integrated dual-wavelength polarization-dependent metalens in SiC is shown Figure 1(a). The design is to simultaneously manipulate the emission from the defect ensemble in the bulk SiC substrate, providing a compact optical interface for multi-channel light collection. The green light corresponds to the orthogonal vertical (V) and horizontal (H) polarized channels at 860 nm, whereas the purple light corresponds to the orthogonal V and H polarized channels at 1240 nm. Figure 1(b) shows the side-view of the monolithic metalens. The device consists of two adjacent metalenses with size of 200 * 200 µm, designed to operate at 860 nm and 1240 nm targeting emissions from Vsi and NV in SiC, respectively. Both metalenses with 1.6 µm height are fabricated on the top surface of a 495 µm-thick bulk SiC substrate and are designed to collect fluorescence emitted from colour centers located on the opposite side of the substrate. The resulting collection half-angle is 21.8°, corresponding to a numerical aperture of NA = n×sin ($\theta$) = 2.6×sin (21.8°) = 0.97, where n is the refractive index of SiC. Figure 1(c) shows the crystal structures of the Vsi and NV centers in 4H-SiC viewed along the [1120] crystallographic direction. Both defects host optically addressable electronic spin states that exhibit long coherence times, making them attractive for quantum sensing and quantum information technologies[29,30]. The optical emission wavelengths of these colour centers depend on their crystallographic configurations and lattice sites. In this work, we consider the zero-phonon-line (ZPL) emissions at 860 nm and 1240 nm, corresponding to the (Vsi) V1 (hexagonal-site) center and the NV (hk) configuration, respectively.

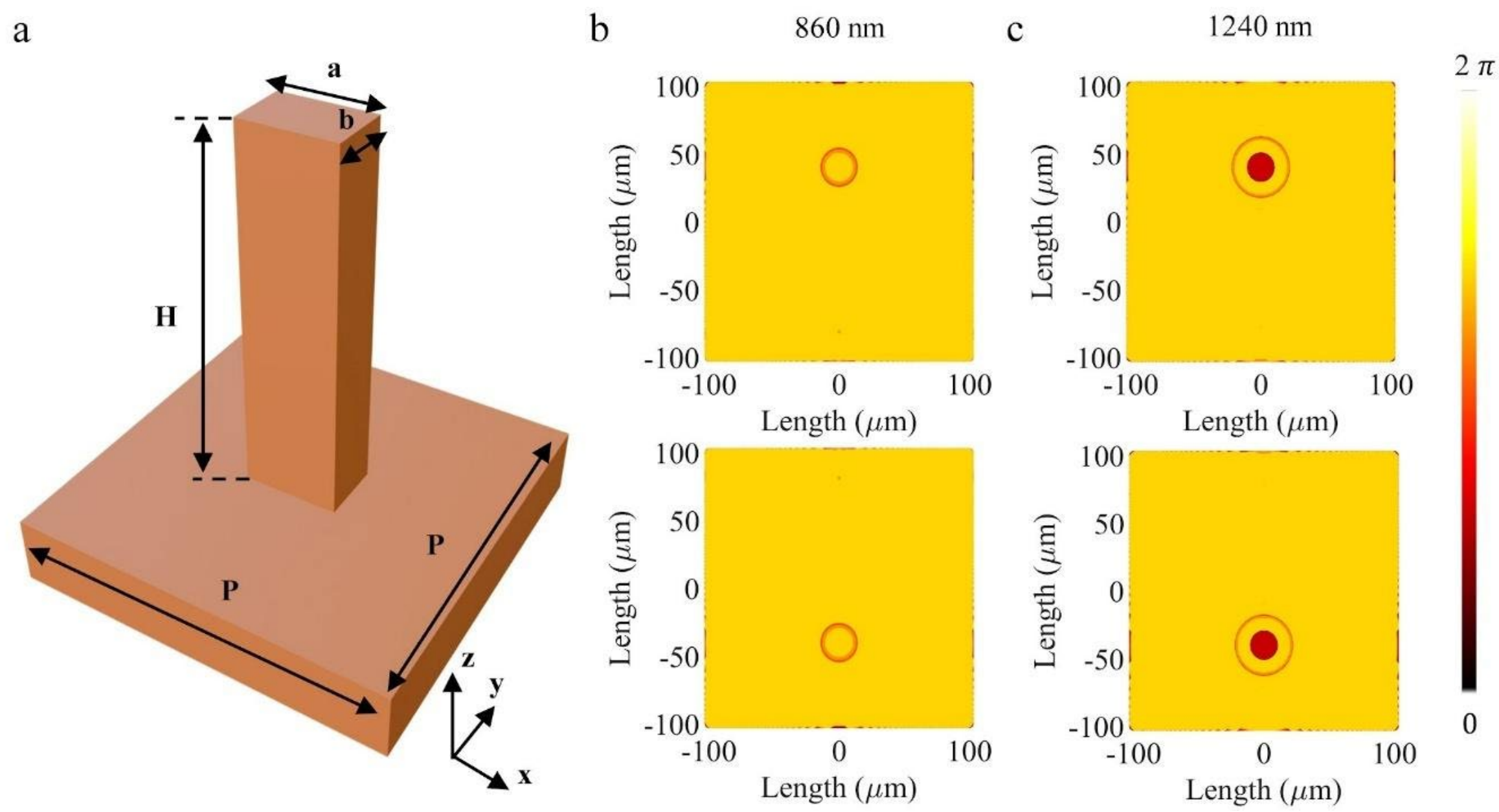

*Figure 2. Metalens simulation. (a)Meta atom parameters used to design the metalens where a and b represent the width and length of the nanopillar and H denotes the height. The lattice period P was chosen based on operating wavelengths of 860 nm or 1240 nm The. (b) Calculated phase distribution for H-polarization (up) and V-polarization (bottom) for 860 nm, respectively. (c) Calculated phase distribution for H-polarization (up) and V-polarization (bottom) for 1240 nm, respectively.*

To realize independent control of the two orthogonal polarization channels, the metalens was designed to impose distinct phase profiles on the V- and H-polarized components, thereby steering them toward opposite directions along the Y-axis at the focal plane. The optical response of the constituent meta-atoms was calculated using rigorous coupled-wave analysis (RCWA) with a square lattice and periodic boundary conditions along the X- and Y-directions. Rectangular cuboid SiC meta-atoms were employed because their two lateral dimensions, a and b, provide independent control over the phase accumulated by the V and H polarized fields, as shown in Figure 2(a). By varying these dimensions, nearly complete 0 to $2\pi$ phase coverage can be obtained for both polarization components while maintaining high transmission.

The meta-atoms and substrate are composed of the same SiC material, with refractive indices of ($n_x = n_y = n_z = 2.645$) at 860 nm and ($n_x = n_y = n_z = 2.619$) at 1240 nm. Different lattice periods were adopted to accommodate the two operating wavelengths, with P = 335 nm for the 860 nm and P = 500 nm for the 1240 nm, while the meta-atom height was fixed at 1.6 μm for both wavelengths to simplify fabrication. Following a full parameter sweep of a and b, eight meta-atom geometries were selected at each wavelength to discretize the required phase response into eight levels. The selected geometries provide high transmission efficiencies exceeding 87% at 860 nm and 86% at 1240 nm (Calculated results in supporting information Figure S1). The target phase profiles were then mapped onto the selected meta-atom libraries to construct the two polarization-manipulating metalenses, with $\pm 40\,\mu m$ separation. The required phase distributions for the H (top) and V (bottom) polarized channels at 860 and 1240 nm are provided and designed to combine emission collection with polarization-dependent beam deflection as shown in figure 2(b) and 2(c), respectively. To verify the resulting optical functionality before full-scale fabrication, finite-difference time-domain (FDTD) simulations were performed on reduced-size metalenses using the discretized geometries. The simulated field distributions and focal-plane intensity profiles presented in supporting information Figure S2 shows clear spatial separation between the two orthogonal polarization channels, confirming the intended polarization-selective response at both operating wavelengths.

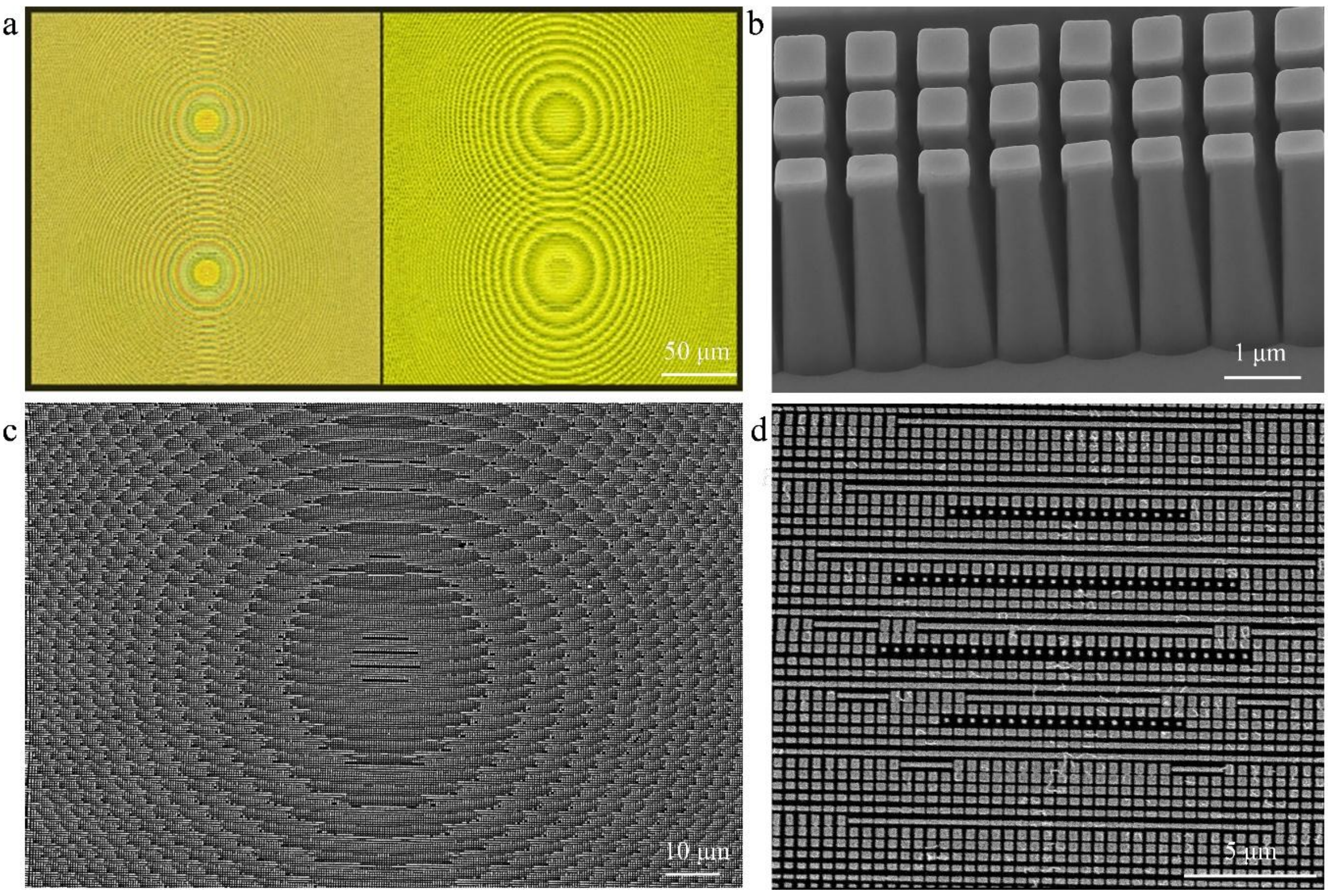

*Figure 3. Metalens fabrication. (a) Optical microscope image of the fabricated monolithic metalens,* where the left part is 860 nm and the right part is 1240 nm design, respectively. *(b) SEM image of the edge region of the metalens. (c) SEM image of the top area of the 1240 nm section. (d) Zoom-in SEM image of the central region of (c).*

The designed metalenses are fabricated using top-down nanofabrication processes, in which the metasurface patterns are defined by electron-beam lithography and transferred on the SiC substrate by lift-off and dry etching (Detail fabrication procedures are described in Figure S3). The optical microscope image of the fabricated monolithic metalens is shown in figure 3(a), where the left part is 860 nm and the right part is 1240 nm design, respectively. The device exhibits excellent uniformity over the entire area without visible fabrication defects indicating the successful fabrication of the large-scale device. The quality of the fabricated metalens was further examined with a scanning electron microscope (SEM) after coating with 5 nm chromium as a conductive layer. Figure 3(b) shows a 45°-tilted SEM image of the edge region of the metalens, showing nanopillars with a height of approximately 1.6 μm. The highly uniform nanopillar array with smooth sidewalls and a nearly vertical profile confirming the etching quality. Figure 3(c) shows a representative top-view SEM image of the 1240 nm section of the metalens, with a magnified view shown in Figure 3(d). The well-defined and uniformly arranged nanostructures over a large area further confirm the high fabrication fidelity and good structural uniformity. (More SEM images of the monolithic metalens are in Figure S4)

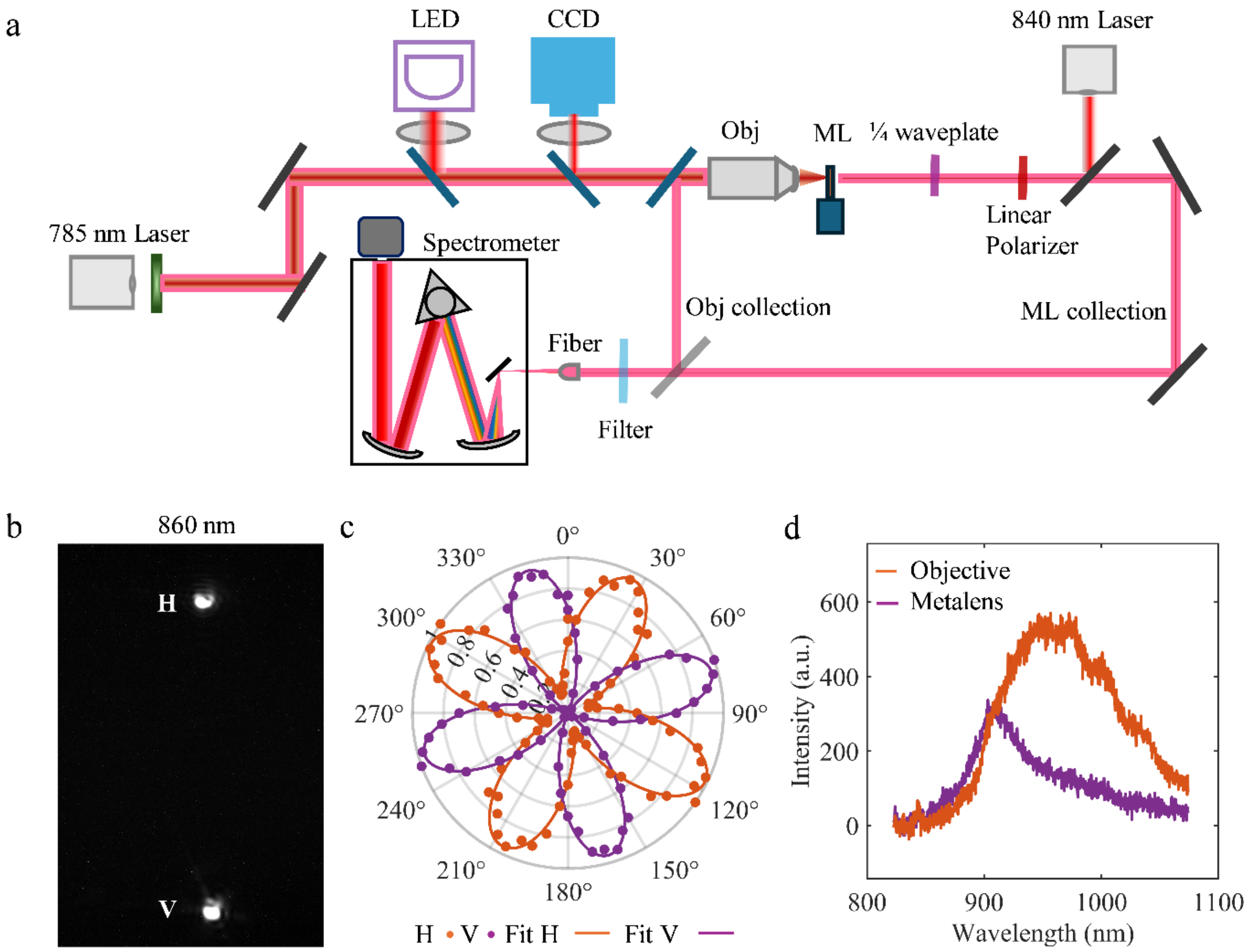

*Figure 4. Characterisation of polarization-selective and efficiency the metalens. (a) Schematics of the set-up. (b) CCD image of the focal plane of the 860 nm design under 840 nm laser illumination, where the up spot represents H-polarized, and bottom spot represents V-polarized. (c) Polarization-resolved intensity of the two focal spots in (b). The scatters are fit to the data showing the two H (orange) and V (purple) polarization states. (d) PL spectra from Vsi in silicon carbide collected through objective (orange) and metalens (purple) at room temperature.*

To characterize the polarization-selective functionality of the metalenses, collimated laser beams at 840 nm and 1300 nm were used to illuminate the metalenses. The resulting focal spots were recorded in transmission using an InGaAs Charge-Coupled Device (CCD, SWIR, 300KMA) camera. The measurement setup for 860 nm design is depicted in figure 4(a). Similar measurements for 1240 nm design are provided in the supporting information Figure S5. As shown in Figure 4(b), the metalens produces two spatially separated focal spots corresponding to the designed H-polarization (upper spot) and V-polarization (lower spot) channels. The spatial separation of V and H polarized light is about 80 μm consistent with our simulation. The polarization selectivity for each wavelength then was characterized using a rotating quarter-wave plate placed after a fixed linear polarizer before the camera. Figure 4(c) shows the corresponding polarization measurement results for the 840 nm illumination focal spots in (a). The orange data points here correspond to H polarization while purple data points are H polarization measurements. The experimental data are fitted by truncated Fourier series. For both wavelengths the visibility of V or H polarizations are > 0.88 confirming selective coupling to orthogonal linear polarization states.

After verifying the polarization-selective operation of the metalens, we next engineer the colour centers on the other side of the SiC wafer using ion implantation. Nitrogen ions were implanted into the SiC substrate with a fluence of $1\times10^{13}$ ions/cm$^2$, to introduce nitrogen dopants and generate vacancies. Subsequently, the sample was annealed to promote diffusion and stabilisation of Vsi. The Vsi was then excited with a 780 nm laser focused by an objective ((NA = 0.9) on the focal point of the metalens. The emission from Vsi was then collected either through the objective or metalens to compare the efficiency of the fabricated metalens (figure 4a). As shown in Figure 4(d) in both cases the broad emission is recorded, while maximum intensity is about half of the one from the objective. From this measurement, the transmission efficiency of the metalens is about 50% for V or H polarization. The relatively lower collection efficiency of the metalens can be mainly attributed to the high refractive index of SiC, which results in substantial Fresnel reflection and limits light extraction at the SiC–air interface due to total internal reflection. It must be noted that there is no polarization selectivity for the collection through the objective arm, hence both V and H polarized lights are collected.

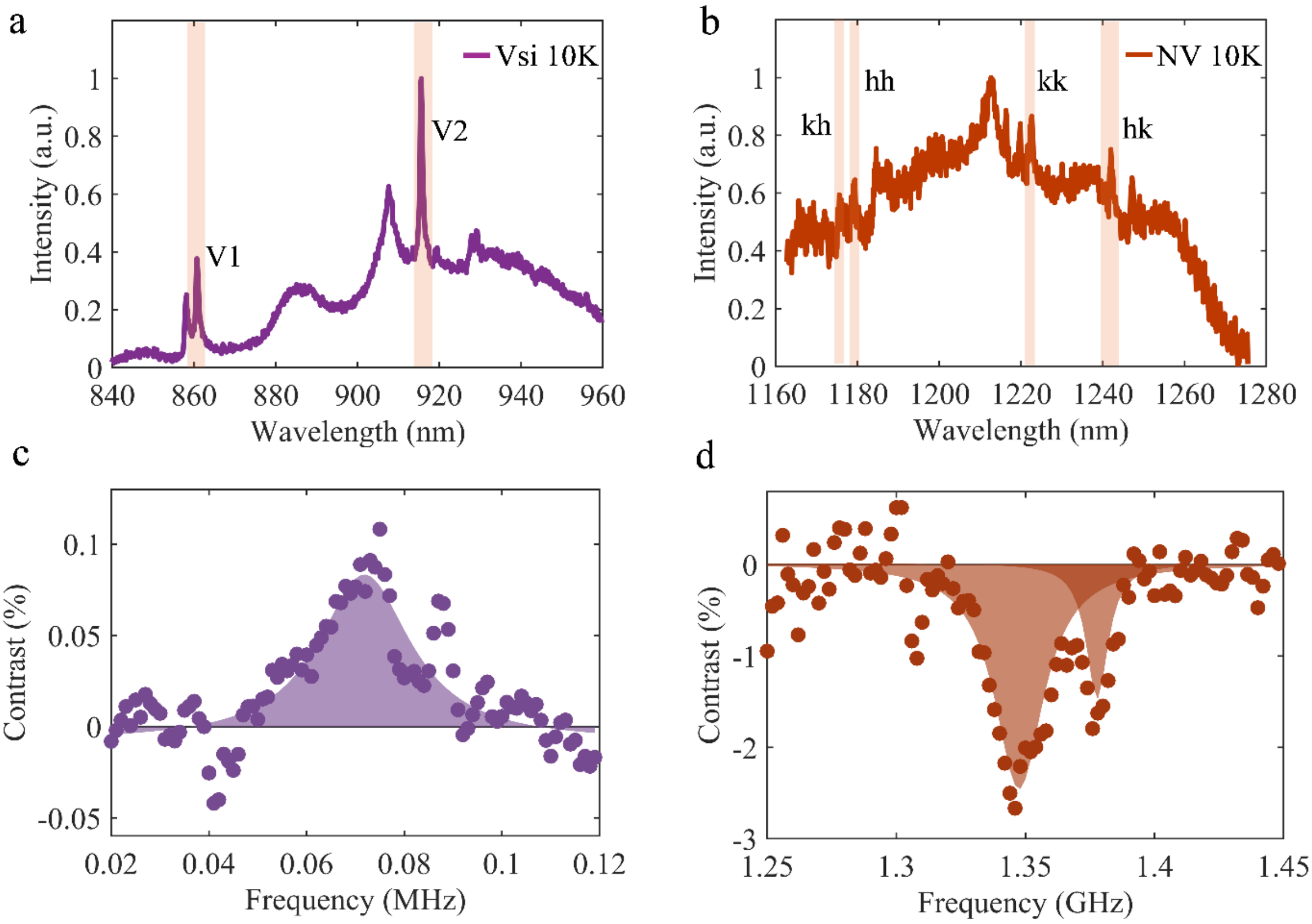

*Figure 5: Spin state readout from the SiC defects through the monolithically integrated metalens. (a) PL spectra collected through the 860 nm design of the metalens under 780 nm laser excitation at 10 K. The orange markers at 861 nm and 917 nm indicate the ZPL of the V1 and V2 silicon-vacancy defects, respectively. (b) PL spectrum collected through the 1240 nm design of the metalens under 980 nm laser excitation at 10 K. The orange markers at 1175.6, 1179.4, 1222.6 and 1242.3 nm indicate the ZPL of the kh, hh, kk, hk nitrogen-vacancy defects, respectively. (c) Room-temperature ODMR spectrum of V1 collected through the metalens. A resonance is observed at ~70 MHz with a contrast of ~0.1%. (d) Room-temperature ODMR spectrum of NV centers collected through the metalens. A resonance at ~1.35 GHz, corresponding to NV defects, is observed with a contrast of ~2.5%. The shaded areas represent the Lorentzian fits to the measured ODMR spectra.*

We further validate the functionality of the metalens by performing PL spectroscopy at 10K where emission spectra from Vsi and NV are better resolved. Figure 5(a) shows the PL spectrum of Vsi collected by the metalens designed for 860 nm. The two sharp zero-phonon lines (ZPLs) at approximately 861 nm and 917 nm are attributed to V1 and V2 configurations of Vsi in 4H-SiC, respectively, as highlighted in the orange markers. Figure 5(b) presents the PL spectrum collected by 1240 nm design from NV centers under 980 nm excitation. In this measurement four types of NV centers with highlight orange markers associated with the kh, hh, kk and hk configurations are clearly observed near 1175.6, 1179.4, 1222.6 and 1242.3 nm, respectively.

These spectral signatures confirm the successful formation of Vsi and NV ensembles in the implanted SiC substrate. While their emission can efficiently be collected by the polarization selective metalens as discussed earlier. This was also confirmed by optically detected magnetic resonance (ODMR) measurement of Vsi and NV. The samples were excited from the backside through the objective lenses, and the emissions from the sample were collected through both the metalens. Figure 5(c) and (d) show the ODMR spectra from Vsi and NV centers, collected through the 860 nm and 1240 nm design of the metalens under zero magnetic field, respectively. The data are smoothed by moving average of three consecutive data points to filter the photon count fluctuation due to the SNSPD vibration. For the Vsi centers, the resonant frequency and contrast measured through the metalens are 71.8 ± 0.6 MHz (72.4 ± 0.4 MHz) and 0.09 ± 0.01%, respectively. For the NV centers, the resonant frequency and contrast measured through the metalens are 1.348 ± 0.001 GHz and 2.5 ± 0.1%, respectively. The resonant frequencies and contrasts are compatible with the values reported in previous studies [31], indicating that the collected PL originates from Vsi and NV centers. The dip at 1.375 GHz corresponds to the resonant frequency of divacancies (PL5) introduced during the ion implantation and annealing process [5]. We also note that these results are comparable to objective lenses collection as shown in supplementary material Figure S6.

## Conclusion

In summary, we have designed and demonstrated monolithic integrated metalenses on SiC substrate for multifunctional quantum state engineering. The designed structures were fabricated with high uniformity and fidelity directly in bulk SiC. The metalens operates at 860 and 1240 nm, enabling light collection and polarization manipulation at both wavelengths. Photoluminescence from silicon-vacancy and nitrogen-vacancy defect ensembles is successfully collected through the metalens. Furthermore, ODMR measurements demonstrate optical spin-state readout through the same device. These results establish a compact optical interface that combines multiwavelength operation, polarization control, and spin readout on a single SiC platform. This work provides a pathway towards multifunctional SiC devices for integrated quantum photonics, quantum sensing, and quantum networking.

## Author Contributions

M.K. and I.A. conceived the project, designed the experiments, and supervised the research. X.Y. H. fabricated the metalens devices, performed the optical characterization, analyzed the experimental data, prepared the figures, and wrote the original manuscript. Z.W.Y. and D.N. contributed to the optical design and numerical simulation of the metalens. K.S. and K.S. performed the optically detected magnetic resonance measurements and assisted with data interpretation. O.C.S contributed to the nanofabrication training. H.H.T. carried out the ion implantation for defect creation. E.W. performed the post-implantation thermal annealing. All authors discussed the results, contributed to the interpretation of the data, reviewed the manuscript, and approved the final version.

**Acknowledgements**

The authors acknowledge financial support from the Australian Research Council (CE200100010, FT220100053, DP250100973) and the Air Force Office of Scientific Research (FA2386-25-1-4044, and the access to nanofabrication facilities through ANFF Sydney Nanoscience Hub, and the UTS Node. We also acknowledge access to NCRIS funded facilities and expertise at the ion-implantation Laboratory (iiLab), a node of the Heavy Ion Accelerator (HIA) Capability at the Australian National University.

**Notes**

There are no conflicts to declare.